# Title: X-ray Nano-imaging of a Heterogeneous Structural Phase Transition in $V_2O_3$

## Authors

Ziming Shao[1], Aileen Luo[1], Eti Barazani[2], Tao Zhou[3], Zhonghou Cai[4], Martin V. Holt[3], Yoav Kalcheim[2, *], Andrej Singer[1, +]

## Affiliations

[1]Department of Materials Science and Engineering, Cornell University; Ithaca, NY 14853, USA

[2]Department of Materials Science and Engineering, Technion, Israel Institute of Technology; Haifa, 3200003, Israel

[3]Center for Nanoscale Materials, Argonne National Laboratory; Lemont, IL 60439, USA

[4]Advanced Photon Source, Argonne National Laboratory; Lemont, IL 60439, USA

*ykalcheim@technion.ac.il

+asinger@cornell.edu

**Abstract**

Controlling the Mott transition through strain engineering is crucial for advancing the development and application of memristive and neuromorphic computing devices. Yet, Mott insulators are heterogeneous due to intrinsic phase boundaries and extrinsic defects, posing significant challenges to fully understanding the impact of local microscopic distortions on the local Mott transition. Addressing these challenges demands structural characterizations at the relevant length scale. Here, using a synchrotron-based scanning X-ray nanoprobe, we studied the real-space structural heterogeneity during the structural phase transition in a $V_2O_3$ thin film. Through temperature-dependent metal-insulator phase coexistence mapping, we report a variation in the local transition temperature of up to 7 K across the film and the presence of the transition hysteresis at the nanoscale. Furthermore, a detailed quantitative analysis demonstrates that the spatial heterogeneity of the transition is closely tied to the tilting of crystallographic planes in the pure insulating phase. Our work highlights the impact of local heterogeneity on the Mott transition and lays the groundwork for future innovations in harnessing strain heterogeneity within Mott systems for the next-generation computational technologies.

**Main Text**

Vanadium sesquioxide ($V_2O_3$) is an archetypal Mott insulator[1–3]. Under ambient conditions, bulk $V_2O_3$ is paramagnetic and metallic (PM). Upon cooling below ~160 K, the material undergoes a discontinuous metal-to-insulator (MIT) phase transition into an antiferromagnetic insulating phase (AFI). Coincidentally, the material undergoes a structural phase transition (SPT) from high-temperature rhombohedral phase (space group $R\bar{3}c$) to low-temperature monoclinic phase (space group $I2/a$), in which the lattice expands in the *ab*-basal plane and contracts along the *c*-axis direction, and the unit cell shears around the *b*-axis[4]. The coupled MIT and SPT temperature is sensitive to hydrostatic pressure[5] and chemical doping[1–3,6,7], which can vary the transition temperature by tens of degrees. The $V_2O_3$ PM phase can also transform into a paramagnetic insulating (PI) phase in Cr-doped $V_2O_3$ at high temperatures, in which the lattice expands in the *ab*-basal plane and contracts along the *c*-axis direction, while retaining the rhombohedral symmetry[1,2].

Epitaxial growth on crystalline substrates enables strain engineering in $V_2O_3$ using the lattice mismatch and differential thermal expansion between the substrate and film[8–13]. This approach has proven effective in tuning the MIT in $V_2O_3$[14–22], which has significant implications for its application in neuromorphic computing[23–26]. Often the transition temperature varies between different regions in the film due to spatial heterogeneity, complicating characterization. Elucidating the role of heterogeneous strain requires observing the structure during the phase transition at a relevant length scale. In this study, we use scanning nano-diffraction to image, with a resolution of 30 nm, the local structural phase transition temperature in a $V_2O_3$ thin film. The analysis of the nano-diffraction data reveals that this local transition temperature varies by 7 K – almost as much as the width of the hysteresis – across the film at a length scale of 100 nm. We find that the local tilts of crystallographic planes best correlate to the local transition temperature. The lattice tilt brings the system closer to the triple point, where all three phases coexist (akin to negative chemical pressure via Cr-doping). Our approach provides simultaneous local measurement and reveals hidden structural characteristics of the phase transition.

The $V_2O_3$ film studied is 100 nm thick and synthesized via sputtering on $(10\bar{1}0)_{R(rhombohedral)}$-oriented sapphire substrates (M-cut)[12,16]. We start the characterization by reciprocal space mapping (RSM) with synchrotron-based parallel X-ray diffraction. At 200 K, the $30\bar{3}0_R$ diffraction peak corresponding to the PM phase dominates the RSM (Fig. 1B(i)). In addition, a much weaker peak is present at a slightly lower momentum transfer $Q_z$ (normal to $(10\bar{1}0)_R$ plane), which is associated with the PI phase: its lower $Q_z$ agrees with the expected expansion of the *ab*-basal plane during the PM to PI transition. In bulk at ambient pressure and room temperature, only the PM phase is stable. The simultaneous observation of both PM and PI phases in the thin film indicates the M-cut sapphire substrate generates strain states spanning across the PI/PM phase boundary akin to negative chemical pressure, as shown as the shaded region in Fig. 1A and reported previously[16].

Upon cooling, the rhombohedral structure transforms into the monoclinic structure by shearing the conventional hexagonal unit cell. Three lower-symmetry monoclinic twins are possible via shearing along the three in-plane axes $[2\bar{1}\bar{1}0]_R$, $[\bar{1}2\bar{1}0]_R$, and $[\bar{1}\bar{1}20]_R$, (see Fig.1 A). The $(30\bar{3}0)_R$ plane in the rhombohedral structure turns into the $(231)_{M(monoclinic)}$, $(402)_M$, and $(2\bar{3}1)_M$ monoclinic planes of the three possible twins[2,4]. At 173 K, three peaks are found in the diffraction pattern (see Fig. 1B(ii)). Consistent with the structural model based on powder diffraction data[2], we associate the two additional Bragg peaks, other than the $30\bar{3}0_R$ peak at top, with the antiferromagnetic insulating (AFI), monoclinic structure: $231_M/2\bar{3}1_M$ peak (bottom right) and $402_M$ peak (middle left). Moreover, we observed that the AFI peaks are shifted along $Q_x$ (parallel to $[0001]_R$), revealing the relative tilt of the monoclinic $(231)_M/(2\bar{3}1)_M$ crystal planes compared to the $(30\bar{3}0)_R$. Upon further cooling to 115 K, well below the PM-AFI transition temperature, the $30\bar{3}0_R$ peak disappears entirely and only the peaks of the AFI phase are observed, indicating the complete transition from the rhombohedral to the monoclinic structure.

To elucidate the ratio of different phases as a function of temperature, we define three distinct regions around the three corresponding diffracted peaks on the detector (white boxes in Fig. 1B(ii)) and then use the relative diffracted intensity of each peak to determine the corresponding phase ratio Φ at various temperatures (see Methods). Figure 1C illustrates that the ratio of PM phase, $\Phi_{PM}$, decreases (increases) upon cooling (heating) with an apparent hysteresis. This confirms the discontinuous nature of the SPT in our $V_2O_3$ thin film. The SPT occurs at approximately 176 K during heating and 168 K during cooling, identified by the point where half of the diffraction intensity is attributed to the PM phase. Concurrently, the ratio of the monoclinic twins $231_M$ and $2\bar{3}1_M$ increases monotonically with the decreasing temperature as expected.

The diffracted intensity of the $402_M$ peak, which is proportional to the phase fraction of that twin, is significantly lower than the other two peaks. This diffracted intensity shows a hysteresis (see inset of Fig.1C), and peaks at the same temperatures as the SPT (176 K during heating and 168 K during cooling), suggesting a significant correlation between the $402_M$ twin phase ratio and phase coexistence between the rhombohedral phase and the $231_M/2\bar{3}1_M$ twins. Given the distinct temperature dependencies of the monoclinic twins, We will refer to the $402_M$ twin as the $AFI_{IT(intermediate\ temperature)}$ phase and to the $231_M$ and $2\bar{3}1_M$ twins as the $AFI_{LT(low\ temperature)}$ phase.

Figure 1D shows the resistance of the same film as a function of temperature. We find an MIT temperature of 173 K (167 K) in the heating (cooling) cycle, similar to the SPT temperatures, indicating the coupling of the SPT and MIT. The MIT observed is wider than in bulk, likely due to the spatial variations in transition temperature[27]. The SPT in the film is also broader than in bulk, suggesting that the SPT transition temperature is also heterogeneous. To understand the origin of the heterogeneity in $V_2O_3$ M-cut film we collected diffraction data similar to those shown in Figure 1B, but now spatially resolved with a scanning X-ray nanoprobe. This method allows us to visualize the spatial distribution of lattice distortions and correlate it with the various observed phases of $V_2O_3$ as they evolve as the temperature changes.

In the nano-diffraction experiment, a monochromatic X-ray beam with photon energy of 10.4 KeV was focused by a Fresnel zone plate to a spot size of 30 nm diameter on the M-cut $V_2O_3$ thin film (Fig. S1). An area detector was used to collect the X-ray diffraction patterns. By raster scanning the X-ray nanoprobe over a 3 μm × 3 μm area of the film with a step size of 30 nm, we acquired a four dimensional dataset of X-ray intensities, I(i, j; k, l), where (i, j) denote the illuminated sample position of the raster scan, and (k, l) denote the coordinates of individual pixels on the area detector. We measured the same area of the film at various temperatures ranging from 115 K to 300 K (see detailed temperature history in Fig. S2).

Figure 2A shows the spatial distribution of the three coexisting phases $\Phi_P(i,j)$ at four temperatures in one heating cycle, where the subscript $P = PM, AFI_{LT}, AFI_{IT}$ indicates the phase. We determined the spatial distribution of different structures by tracking their X-ray diffraction intensity, similar to the dark field scanning transmission electron microscopy and our analysis in Figure 1. At the lowest temperature, most of the area is occupied by the $AFI_{LT}$ phase, consistent with the parallel beam measurements (see Fig. 1). With increasing temperature, the maps show a phase transition from the $AFI_{LT}$ to the PM structural phase. The spatial distributions of the $AFI_{LT}$ and PM phases are anticorrelated, revealing phase separation, as expected in a discontinuous phase transition. The morphology of the domains of different phases during the phase coexistence shows no clear alignment with respect to crystallographic axes. We observe puddles of the PM phase that nucleate and grow inside of the $AFI_{LT}$ phase. This morphology differs from the ordered domains observed using PEEM[28,29] in thicker $V_2O_3$ grown on $(0001)_R$-oriented or using nano-IR imaging in thicker films grown on $(01\bar{1}2)_R$-oriented sapphire[30]. The domain morphology resembles the one observed using X-ray nano-diffraction in thinner $V_2O_3$ films grown $(01\bar{1}2)_R$-oriented sapphire[31]. The relative ratio of the $AFI_{IT}$ phase is low at 115 K and increases slightly during the heating without significantly changing its spatial distribution.

Figure 2B overlays the spatial distribution of all the three phases at 176 K. It reveals that the $AFI_{IT}$ phase domains are predominantly situated near the junctions between PM and $AFI_{LT}$ domains (see regions highlighted by red circles in Fig. 2B), although not consistently at every junction. The correlation of this spatial distribution with reciprocal space maps in Figure 1—where the $AFI_{IT}$ phase exhibits its strongest diffraction signal in the co-presence of both PM and $AFI_{LT}$ phases— indicates that the $AFI_{IT}$ phase likely resides at the $AFI_{LT}$-PM junctions. The out-of-plane lattice spacing of the $AFI_{IT}$ phase ($(402)_M$ planes), is between those of $AFI_{LT}$ phase ($231_M/2\bar{3}1_M$ planes) and PM phase ($(30\bar{3}0)_R$ planes). This suggests that in the epitaxial thin film where the in-plane strain is constrained by the substrate, the $AFI_{IT}$ phase relaxes the out-of-plane strain at the junctions between $AFI_{LT}$ and PM phases. Nevertheless, the absence of strong $AFI_{IT}$ signal at some $AFI_{LT}$-PM junctions implies additional factors requiring further investigation. For example, the porosity in similarly synthesized films[14] could result in strain relaxation of $AFI_{LT}$ or PM phases at voids without the presence of the $AFI_{IT}$ phase.

Figure 2C illustrates the spatial distribution of the PM phase $\Phi_{PM}(i,j)$, at 200 K and 300 K, measured during the same heating cycle as the scans presented in Figure 2A. Consistent with the observations from the unfocused beam measurement, most of the film's volume is dominated by the PM phase, with the remaining portion occupied by the PI phase. Notably, the phase ratio of the PI phase is greater at 300 K than at 200 K, although the overall morphology of the phases remains similar. Intriguingly, the morphology of the PM/PI phase coexistence at higher temperatures bears resemblance to that of the PM/AFI phase coexistence observed at lower temperatures (see regions highlighted by the white boxes in Figure 2A and 2C). A Pearson correlation coefficient of 0.45 is observed between the boxed regions at temperatures of 176 K and 300 K. This morphological correlation suggests a direct transition pathway from the AFI to the PI phase. Moreover, the morphology observed in our $V_2O_3$ film is reproducible after heating to room temperature and cooling below the SPT temperature (see Figure S4). Similar reproducibility has been observed in $V_2O_3$ films using optical microscopy and in X-ray PEEM experiments[27–29] and in $VO_2$ films in pump probe experiments[32], where the morphology is stable to repeated transitions excited by laser exposures. Also, we observe similar phase morphology by comparing the spatial distribution of the structural phases during cooling and heating (see Fig. S5). This similarity suggests the structural transition hysteresis is present locally at the nanoscale.

We use the temperature series shown in Figure 2A to classify different regions of the film by their temperature behavior. To this end, we introduce a local parameter $\bar{\Phi}(i,j)$ as:

$$\bar{\Phi}(i,j) = \sqrt{\sum_{n=1}^{N}[\Phi_{PM}(i,j;T_n)]^2} \,,$$

where $\Phi_{PM}(i,j;T_n)$ is the PM phase ratio calculated from the X-ray data collected at sample position $(i,j)$ and temperature $T_n$, and the sum is over all N measured temperatures. This projection of the 5D dataset describes the local propensity for the PM phase. Figure 3A shows $\bar{\Phi}(i,j)$ calculated from the four temperature scans showed in Figure 2A. To study the spatial heterogeneity, we categorize all measured sample positions, $(i,j)$, into five regions (R0-R4) by grouping them into bins based on the histogram of all $\bar{\Phi}$ values as shown in Figure 3B (similar results emerge with more bins, see Fig. S8). We then average the diffraction profile for each region and temperature. This classification procedure allows us to extract diffraction profiles from areas with distinct transition behaviors. These areas are separated by tens to hundreds of nanometers (delineated by contour lines in Fig. 3A).

From these categorized diffraction data, we calculate the relative ratio of the coexisting structural phases as a function of the temperature, similar to the analysis in Figure 1, but for the nanoscale regions defined (see Fig. 3C). As expected, the regions with a higher $\bar{\Phi}$ values transition first into the PM phase upon heating. To determine the local transition temperature, we fit the temperature dependence of the PM phase fraction in each region with a logistic function, $\frac{1}{e^{-A(T-T_{SPT})}+1}$, where parameter A is determined by fitting the curve in Figure 1C up to 175 K. We find that the transition temperature, $T_{SPT}$, increases from 176 K in R4 to 183 K in R0 (see Fig. S6).

At 115K, the averaged diffraction profiles across regions show a Bragg peak shift perpendicular to $Q_z$, indicating varying tilt levels in the $(231)_M$ and $(2\bar{3}1)_M$ lattice planes away from the film normal $[0001]_R$ (see Fig. S7). Mapping these shifts with the calculated transition temperatures, Figure 3 reports PM-AFI phase boundary relative to temperature and peak shift. The tilting of these lattice planes correlates with the local transition temperature, similar to the pressure effects in bulk $V_2O_3$ (see Fig. 1A). The regions with a lower AFI-PM transition temperature show a smaller relative PI phase ratio at 200 K and 300 K, both above the AFI-PM SPT. This indicates a higher PM-PI transition temperature in regions with a lower AFI-PM transition temperature and underscores the similarity between the crystal plane tilts in thin films and the pressure effects in bulk $V_2O_3$.

By plotting the profiles of the $AFI_{LT}$ peak at various temperatures in Figure 4A, we observe an asymmetric variation in peak intensity, with a more significant decrease on the low $Q_x$ side during heating. This results in a shift of the peak center of mass towards the $+Q_x$ direction, as indicated by the triangles in the bottom of the figure. To track the volume variation of the monoclinic structures with different level of crystalline plane tilting, we divide the peak into three equal intervals (LT1-LT3) based on the $Q_x$ range and plot their spatial distribution at various temperatures during heating, as shown in Figure 4B. The significant decrease in intensity in LT1 and LT2 maps indicates the shrinking monoclinic volume due to SPT. Meanwhile, the intensity in LT3 remains largely unchanged upon heating to 176K. This demonstrates that the monoclinic structures with crystalline plane tilting contributing to the LT3 have higher transition temperatures, corroborating our previous finding on the correlation between SPT temperature and local tilts in the $(231)_M$ and $(2\bar{3}1)_M$ lattice planes. Furthermore, the maps in each column (the same $Q_x$ interval) show high correlation, indicating the local monoclinic lattice plane tilting is not affected by temperature variation or SPT.

In our study, we investigated the impact of structural heterogeneity on the Mott transition in $V_2O_3$ thin films grown on $(10\bar{1}0)$-oriented sapphire substrates using a synchrotron-based scanning X-ray nanoprobe. We made three observations: (1) While in bulk $V_2O_3$ all three monoclinic twins are present equivalently, in the thin film grown on the M-cut $Al_2O_3$, two twins dominate. The other twin structure reaches a maximum volume fraction during phase coexistence and diminishes upon further cooling. This behavior suggests a mechanism where the minimization of strain energy at the junctions between coexisting phases dictates the formation of the preferred twin structure. (2) Through quantitative spatial analysis, we revealed a variation of local transition temperature of 7 K and the presence of the transition hysteresis locally at the nanoscale. (3) The spatially heterogeneous transition temperature correlates with the tilt of the monoclinic crystallographic planes, which is insensitive to temperature variation and SPT. These insights affirm the crucial impact of structural heterogeneity on Mott transition and shed light on precise phase transition manipulation through microscopic strain engineering.

Going beyond the two-dimensional spatially resolved structural analysis presented here, we anticipate that future progress in understanding heterogeneity in Mott insulators could benefit from three key methodological advancements. First, the advancing from two-dimensional to three-dimensional imaging will allow to better resolve the orientation and structure of interface between the coexisting phases and the twin structures[33,34]. Second, a large-volume reciprocal space mapping, including multiple Bragg peaks[35], applied to thin films will provide more constraints simultaneously than those provided by a single diffraction peak. We expect this to enable the development of a robust structural model correlating the observed lattice tilting and strain. Third, the simultaneous spatially resolved observation of the metal-insulator transitions and structural phase transitions is promising to unravel the subtle interplay of electronic, spin, orbital and structural degrees of freedom in these complex materials.

## Methods

### V$_2$O$_3$ Thin Film Preparation

100 nm V$_2$O$_3$ films were grown on sapphire (Al$_2$O$_3$) substrates of ($10\bar{1}0$)-orientations by RF magnetron sputtering as described previously in Trastoy, *et al.*[12] and Barazani, *et al*[16].

### X-ray Scanning Nano-Probe

The X-ray scanning nano-probe measurements were conducted at beamline 26-ID at the Advanced Photon Source with the incident photon energy of 10.4 KeV. A liquid-nitrogen cooled Si (111) double crystal monochromator (DCM) was used to achieve high energy resolution (ΔE/E = 1.7 x 10$^{-4}$) and tune the x-ray energy to 10.4 keV. A Fresnel zone plate with outside diameter of 133 µm and outermost zone width of 24 nm combined with order sorting aperture was used to focus the collimated x-ray beam to 30 nm (FWHM) diameter with flux of ~10$^9$ photons/s. The nanoprobe scans were taken with ~30 nm step size.

### Data Analysis

Regions of Interest (ROIs) are selected around the diffraction peaks are applied in the data analysis. As shown in Figure 1B(ii), the ROI on top corresponds for the $30\bar{3}0_R$ peak of the PM phase. The ROI at middle left is for the 402$_M$ peak of the AFI phase. And the ROI at the bottom right is for the aggregation of 231$_M$/2-31$_M$ peak of the AFI phase at low temperatures and $30\bar{3}0_R$ of the PI phase at high temperatures. Because of the strong overlapping of the 231$_M$/2-31$_M$ peak and PI peak, the transition between the PI and AFI phases is indistinguishable in our diffraction measurement. Based on the canonical phase diagram of V$_2$O$_3$, we speculate the PI/AFI transition temperature at ~180K[3].

The relative phase ratios $\Phi_P$ in this study, where $P = PM, AFI_{LT}, AFI_{IT}$, are calculated by comparing the integrated diffraction intensities of the corresponding phases. The total intensity of the RSM (sum of all peaks) varies by only 10% among all the temperatures (see Figure S3). Considering the minor intensity variation, we assume the structural factor to be identical for both the rhombohedral and monoclinic phases.

The relative phase ratio $\Phi_P$ of parallel beam measurement shown in Figure 1C is defined as:

$$\Phi_P = \sum_{(k,l) \in ROI_P} I(k,l) / \sum_{(k,l)} I(k,l),$$

where (k, l) denotes the coordinates of individual pixel on the area detector.

In the nanoprobe measurements, two additional coordinates (i, j) are introduced to represent the sample position illuminated by the focused X-ray beam in real space. Spatial distribution of the phases $\Phi_P(i,j)$ shown in Figure 2A is defined as:

$$\Phi_P(i,j) = \sum_{(k,l) \in ROI_P} I(i,j;k,l) / \sum_{(k,l)} I(i,j;k,l).$$

The $\Phi_p^R(T)$ shown in Figure 3C is defined as:

$$\Phi_p^R(T) = \frac{1}{|R|} \sum_{(i,j) \in R} \left( \sum_{(k,l) \in ROI_p} I_T(i,j;k,l) / \sum_{(k,l)} I_T(i,j;k,l) \right),$$

where $p \in \{AFI, PM\}, R \in \{R0, R1, R2, R3, R4\}, T \in \{115\,K, 170\,K, 173\,K, 176\,K\}$.

## Acknowledgements

Z.S., E.B., Y.K., and A.S. acknowledge support from U.S. - Israel Binational Science Foundation (BSF), under Grant No. 2020337. Work performed at the Center for Nanoscale Materials and Advanced Photon Source, both U.S. Department of Energy Office of Science User Facilities, was supported by the U.S. DOE, Office of Basic Energy Sciences, under Contract No. DE-AC02-06CH11357. A.L. acknowledges a graduate research fellowship through the National Science Foundation (No. DGE-2139899).

**References**

1.	McWhan, D. B., Rice, T. M. & Remeika, J. P. Mott Transition in Cr-Doped V 2 O 3. *Phys. Rev. Lett.* **23**, 1384–1387 (1969).

2.	McWhan, D. B. & Remeika, J. P. Metal-Insulator Transition in ( V 1 − x Cr x ) 2 O 3. *Phys. Rev. B* **2**, 3734–3750 (1970).

3.	Rice, T. M. & McWhan, D. B. Metal-insulator Transition in Transition Metal Oxides. *IBM J. Res. Dev.* **14**, 251–257 (1970).

4.	Dernier, P. D. & Marezio, M. Crystal Structure of the Low-Temperature Antiferromagnetic Phase of V 2 O 3. *Phys. Rev. B* **2**, 3771–3776 (1970).

5.	Limelette, P. *et al.* Universality and Critical Behavior at the Mott Transition. *Science* **302**, 89–92 (2003).

6.	McWhan, D. B., Menth, A., Remeika, J. P., Brinkman, W. F. & Rice, T. M. Metal-Insulator Transitions in Pure and Doped V 2 O 3. *Phys. Rev. B* **7**, 1920–1931 (1973).

7.	Kuwamoto, H., Honig, J. M. & Appel, J. Electrical properties of the (V1-xCrx)2O3 system. *Phys. Rev. B* **22**, 2626–2636 (1980).

8.	Schuler, H., Klimm, S., Weissmann, G., Renner, C. & Horn, S. Influence of strain on the electronic properties of epitaxial V2O3 thin films. *Thin Solid Films* **299**, 119–124 (1997).

9.	Yonezawa, S., Muraoka, Y., Ueda, Y. & Hiroi, Z. Epitaxial strain effects on the metal–insulator transition in V2O3 thin films. *Solid State Commun.* **129**, 245–248 (2004).

10.	Dillemans, L. *et al.* Evidence of the metal-insulator transition in ultrathin unstrained V2O3 thin films. *Appl. Phys. Lett.* **104**, 071902 (2014).

11.	Majid, S. S. *et al.* Stabilization of metallic phase in V2O3 thin film. *Appl. Phys. Lett.* **110**, 173101 (2017).

12.	Trastoy, J., Kalcheim, Y., Del Valle, J., Valmianski, I. & Schuller, I. K. Enhanced metal–insulator transition in V2O3 by thermal quenching after growth. *J. Mater. Sci.* **53**, 9131–9137 (2018).

13.	Wickramaratne, D., Bernstein, N. & Mazin, I. I. Impact of biaxial and uniaxial strain on ${\mathrm{V}}_{2}{\mathrm{O}}_{3}$. *Phys. Rev. B* **100**, 205204 (2019).

14. Valmianski, I., Ramirez, J. G., Urban, C., Batlle, X. & Schuller, I. K. Deviation from bulk in the pressure-temperature phase diagram of V 2 O 3 thin films. *Phys. Rev. B* **95**, 155132 (2017).

15. Metcalf, P. A. *et al.* Electrical, structural, and optical properties of Cr-doped and non-stoichiometric V2O3 thin films. *Thin Solid Films* **515**, 3421–3425 (2007).

16. Barazani, E. *et al.* Positive and Negative Pressure Regimes in Anisotropically Strained V2O3 Films. *Adv. Funct. Mater.* **n/a**, 2211801 (2023).

17. Kalcheim, Y. *et al.* Structural Manipulation of Phase Transitions by Self-Induced Strain in Geometrically Confined Thin Films. *Adv. Funct. Mater.* **30**, 2005939 (2020).

18. Kalcheim, Y. *et al.* Robust Coupling between Structural and Electronic Transitions in a Mott Material. *Phys. Rev. Lett.* **122**, 057601 (2019).

19. Homm, P., Menghini, M., Seo, J. W., Peters, S. & Locquet, J.-P. Room temperature Mott metal–insulator transition in V2O3 compounds induced via strain-engineering. *APL Mater.* **9**, 021116 (2021).

20. Querré, M. *et al.* Metal–insulator transitions in (V1-xCrx)2O3 thin films deposited by reactive direct current magnetron co-sputtering. *Thin Solid Films* **617**, 56–62 (2016).

21. Alyabyeva, N. *et al.* Metal-insulator transition in V2O3 thin film caused by tip-induced strain. *Appl. Phys. Lett.* **113**, 241603 (2018).

22. Brockman, J., Samant, M. G., Roche, K. P. & Parkin, S. S. P. Substrate-induced disorder in V2O3 thin films grown on annealed c-plane sapphire substrates. *Appl. Phys. Lett.* **101**, 051606 (2012).

23. del Valle, J., Ramírez, J. G., Rozenberg, M. J. & Schuller, I. K. Challenges in materials and devices for resistive-switching-based neuromorphic computing. *J. Appl. Phys.* **124**, 211101 (2018).

24. Del Valle, J. *et al.* Subthreshold firing in Mott nanodevices. *Nature* **569**, 388–392 (2019).

25. Salev, P., del Valle, J., Kalcheim, Y. & Schuller, I. K. Giant nonvolatile resistive switching in a Mott oxide and ferroelectric hybrid. *Proc. Natl. Acad. Sci.* **116**, 8798–8802 (2019).

26. Kalcheim, Y. *et al.* Non-thermal resistive switching in Mott insulator nanowires. *Nat. Commun.* **11**, 2985 (2020).

27. Lange, M. *et al.* Imaging of Electrothermal Filament Formation in a Mott Insulator. *Phys. Rev. Appl.* **16**, 054027 (2021).

28. Ronchi, A. *et al.* Early-stage dynamics of metallic droplets embedded in the nanotextured Mott insulating phase of V 2 O 3. *Phys. Rev. B* **100**, 075111 (2019).

29. Ronchi, A. *et al.* Nanoscale self-organization and metastable non-thermal metallicity in Mott insulators. *Nat. Commun.* **13**, 3730 (2022).

30. McLeod, A. S. *et al.* Nanotextured phase coexistence in the correlated insulator V2O3. *Nat. Phys.* **13**, 80–86 (2017).

31. Singer, A. *et al.* Nonequilibrium Phase Precursors during a Photoexcited Insulator-to-Metal Transition in V 2 O 3. *Phys. Rev. Lett.* **120**, 207601 (2018).

32. Johnson, A. S. *et al.* Ultrafast X-ray imaging of the light-induced phase transition in VO2. *Nat. Phys.* **19**, 215–220 (2023).

33. Pfeiffer, F. X-ray ptychography. *Nat. Photonics* **12**, 9–17 (2018).

34. Holler, M. *et al.* High-resolution non-destructive three-dimensional imaging of integrated circuits. *Nature* **543**, 402–406 (2017).

35. Krogstad, M. J. *et al.* The relation of local order to material properties in relaxor ferroelectrics. *Nat. Mater.* **17**, 718–724 (2018).

36. Last, B. J. & Thouless, D. J. Percolation Theory and Electrical Conductivity. *Phys. Rev. Lett.* **27**, 1719–1721 (1971).

# Figures

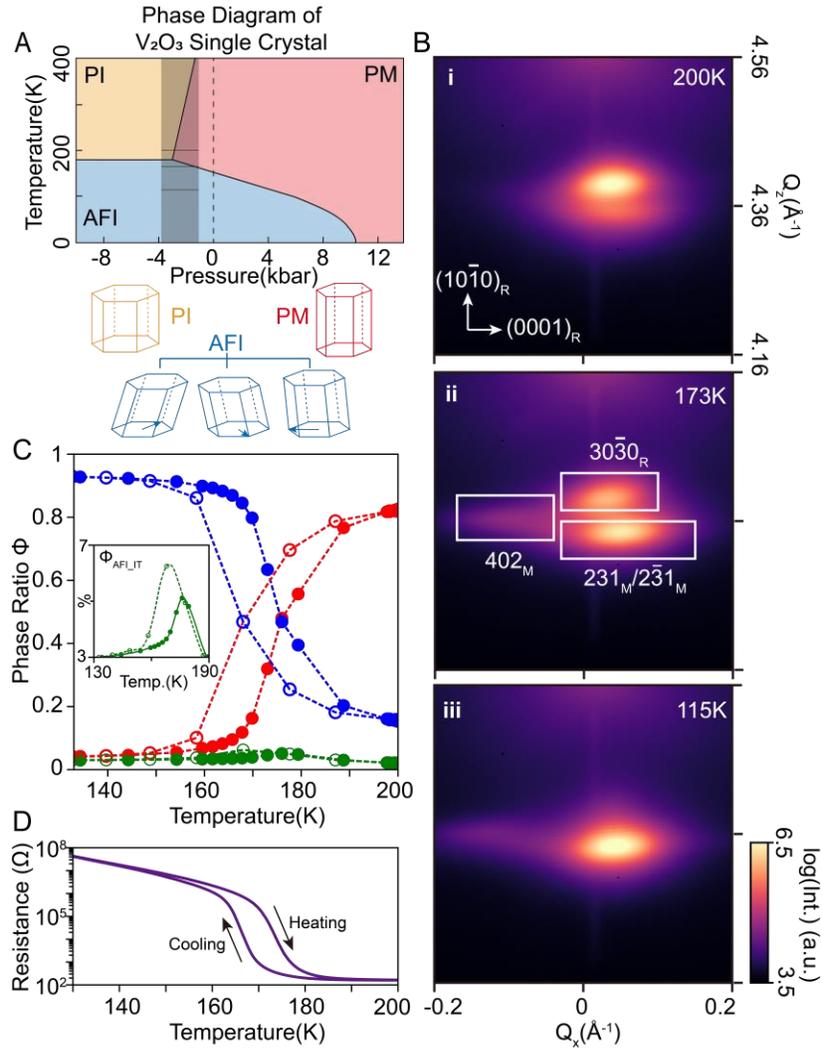

**Fig. 1. (A)** The canonical phase diagram of $V_2O_3$ reproduced from ref.[6] shows three major phases that exhibit distinct magnetic orders and electrical behaviors, including Paramagnetic Metallic (PM), Paramagnetic Insulating (PI), and Antiferromagnetic Insulating (AFI). **(B)** The 3D RSM around the $V_2O_3$ $30\bar{3}0_R$ peak projected along $[\bar{1}2\bar{1}0]_R$ direction at various temperatures. **(C)** The phase ratios in $V_2O_3$ thin film estimated by the relative diffracted intensity. The solid and open symbols are for the heating and cooling cycles, respectively. And the colors correspond to the integrated intensity in the three ROIs shown in panel B (ii); the red, blue and green colors represent for the PM phase, the aggregation of PI (above 180 K) and $AFI_{LT}$ (below 180 K), and $AFI_{IT}$ phase. **(D)** The resistance of the $V_2O_3$ thin film as a function of temperatures.

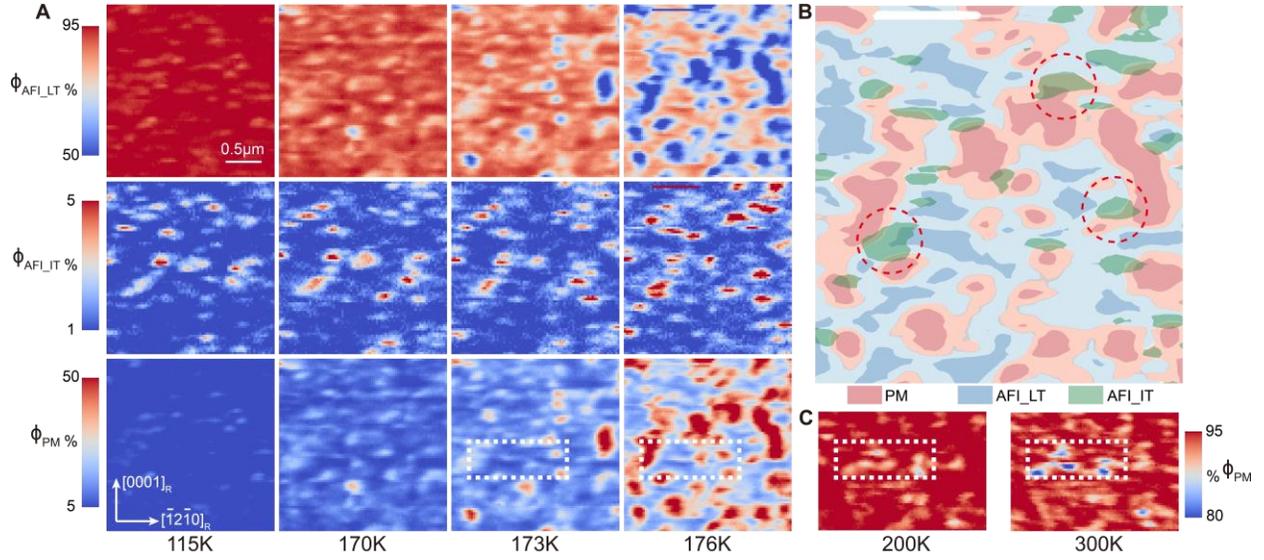

**Fig. 2. (A)** Spatial distribution of the competing phases in V$_2$O$_3$, $\Phi_P(i,j)$, where $P \in [PM, AFI_{LT}, AFI_{IT}]$ at temperatures across its insulator-to-metal (AFI-to-PM) transition. The Horizontal and vertical axes of the map correspond to the $[\bar{1}2\bar{1}0]_R$ and $[0001]_R$ axes of the rhombohedral lattice, respectively. **(B)** Superimposed contour maps of the spatial distributions of PM, AFI$_{LT}$ and AFI$_{IT}$ phases. The AFI$_{IT}$ domains appear mostly around the boundary of the PM and AFI$_{LT}$ phases. The contour levels used are $[avg(\Phi_P), avg(\Phi_P) + std(\Phi_P)]$ for $P \in [PM, AFI_{LT}]$, and $[avg(\Phi_P) + std(\Phi_P)]$ for $P \in [AFI_{IT}]$. **(C)** Spatial distribution of the PM phase $\Phi_{PM}(i,j)$ at 200 K and 300 K. The white dash boxes are plotted for reference, indicating the same sample area for scans at different temperatures.

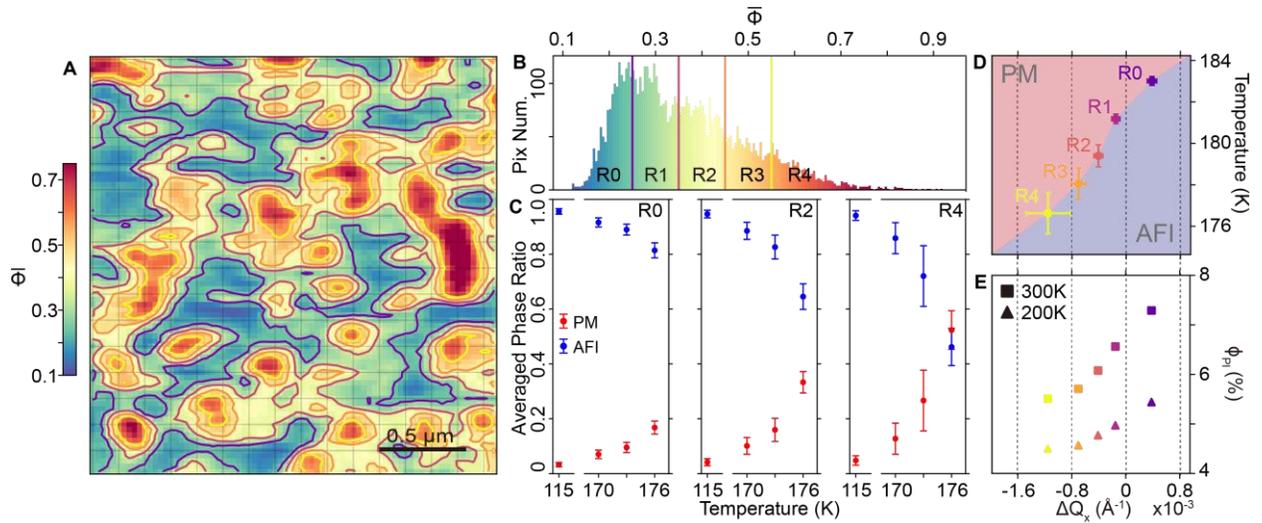

**Fig. 3. (A)** Spatial distribution of the structural transition propensity $\bar{\Phi}$. **(B)** Histogram illustrates the distribution of measured real space locations with respect to structural transition propensity $\bar{\Phi}$. The horizontal axis represents the $\bar{\Phi}$ values, and the vertical axis of the histogram represents the area or number of measured positions with the corresponding range of the $\bar{\Phi}$ values. The sample positions are divided into five intervals with equidistant $\bar{\Phi}$ values. The contour lines in panel (A) share the color with the vertical lines in the histogram, delineating the regions of intervals with different $\bar{\Phi}$. **(C)** The averaged phase ratio as a function of temperatures for different regions of intervals $\Phi_p^{Ri}(T)$. The region of interval with larger $\bar{\Phi}$ values have a lower PM-AFI transition temperature. **(D)** The averaged transition temperature and the relative peak shift along $Q_x$ for regions R0 to R4, which concurrently determine the phase boundary of the PM and AFI phases. **(E)** The relative PI phase ratio for the defined regions at 200 K (triangle) and 300 K (square). The same horizontal axis is shared by the vertically aligned panels 3D and 3E.

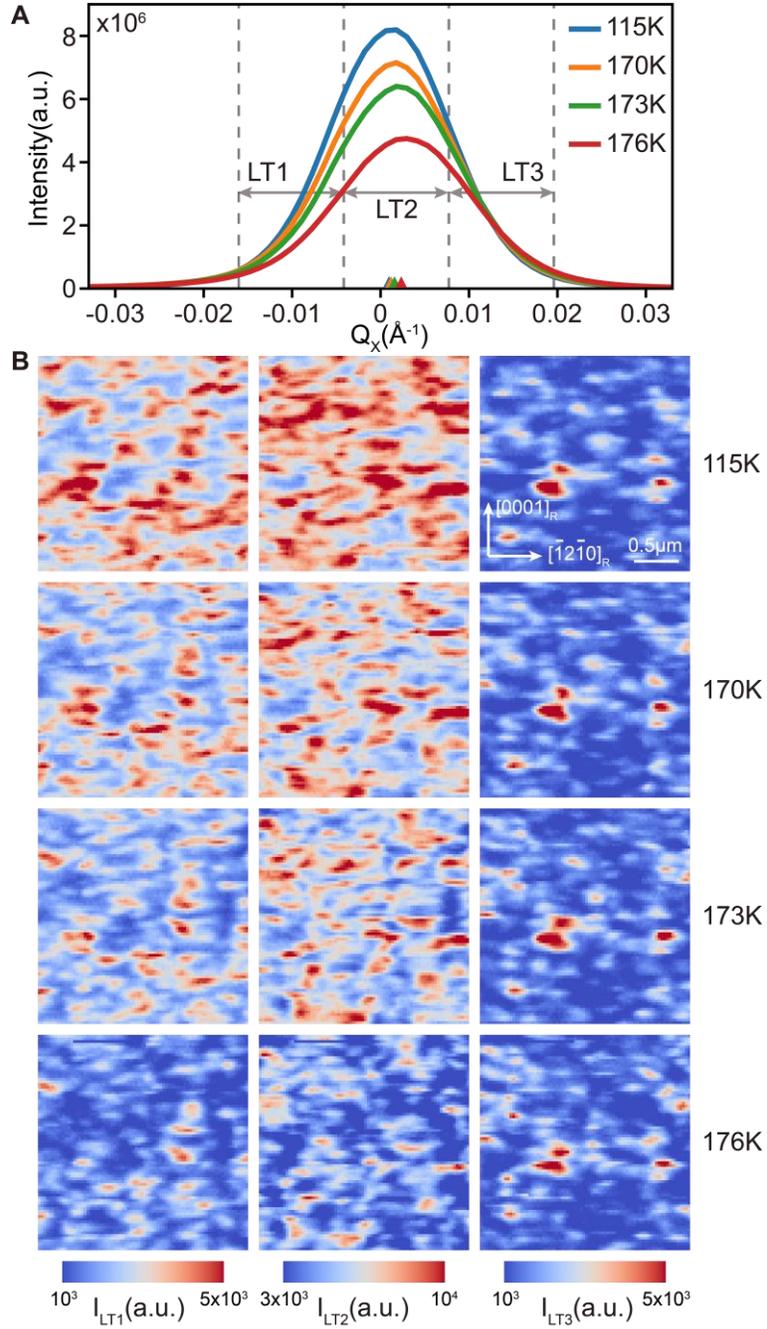

**Fig. 4. (A)** The projections of the AFI$_{LT}$ peak along Q$_x$ direction under different temperatures in the heating cycle. The peak is further separated into three equidistant intervals [$LT1, LT2, LT3$]. **(B)** Spatial distribution of the AFI$_{LT}$ phase with contribution to different Q$_x$ intervals (different level of crystallographic plane tilting), $I_P(i,j)$, where $P \in [LT1, LT2, LT3]$ under different temperatures. The regions with tilting in the positive Q$_x$ direction undergo a transition at higher temperature, consistent with Figure 3.

## Supplementary Figures

**Figure S1**

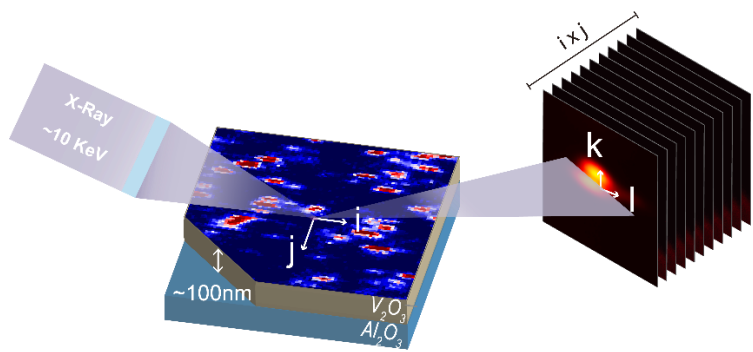

Schematic illustration of experimental setup for scanning X-ray nanoprobe diffraction measurements on a 100 nm $V_2O_3$ thin film grown on M-cut ($10\bar{1}0$-oriented) sapphire substrate.

**Figure S2**

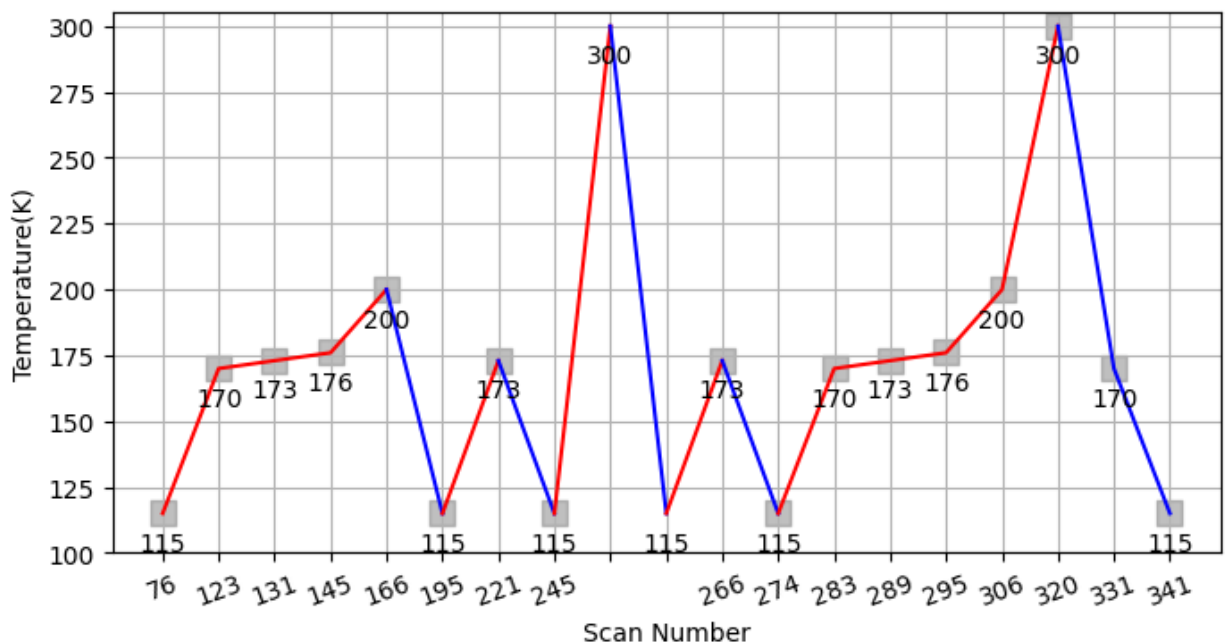

Temperature history of the X-ray nanoprobe experiment. The red and blue curves represent the heating and cooling cycles, respectively. X-ray nanoprobe scans are performed at the temperatures highlighted by grey boxes. The data for Figure 2 in the main text comes from the last heating cycle (Scan 274 - Scan 320).

**Figure S3**

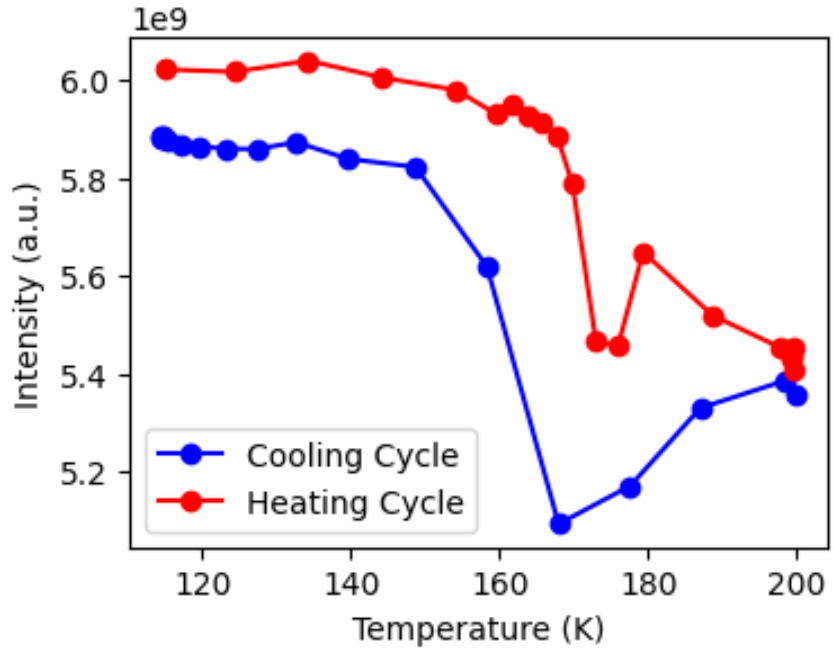

The integrated intensity of the Bragg peak as a function of temperatures. The integrated intensity is higher at the low temperatures when compared to the high temperatures. Notably, an intensity dip is observed at intermediate temperatures corresponding to the structural phase transition in both heating and cooling branch. The observation cannot be readily explained by the different structural factors of the phases, which will lead to a monotonic intensity variation without the intensity dip. Instead, we speculate the measured intensity variation may correlate to the phases' coexistence in the $V_2O_3$ thin film, specifically the coexistence of AFI and PM phases during SPT, or the coexistence of PM and PI phases at higher temperatures. The observed intensity variations may be attributed to the decreased crystallinity at the interfaces between different phases, which impacts the diffraction intensity. Quantification of inhomogeneous crystallinity is beyond the scope of our study. In the present work, the relative phase ratios are determined by comparing the integrated diffraction intensities of the corresponding phases.

**Figure S4**

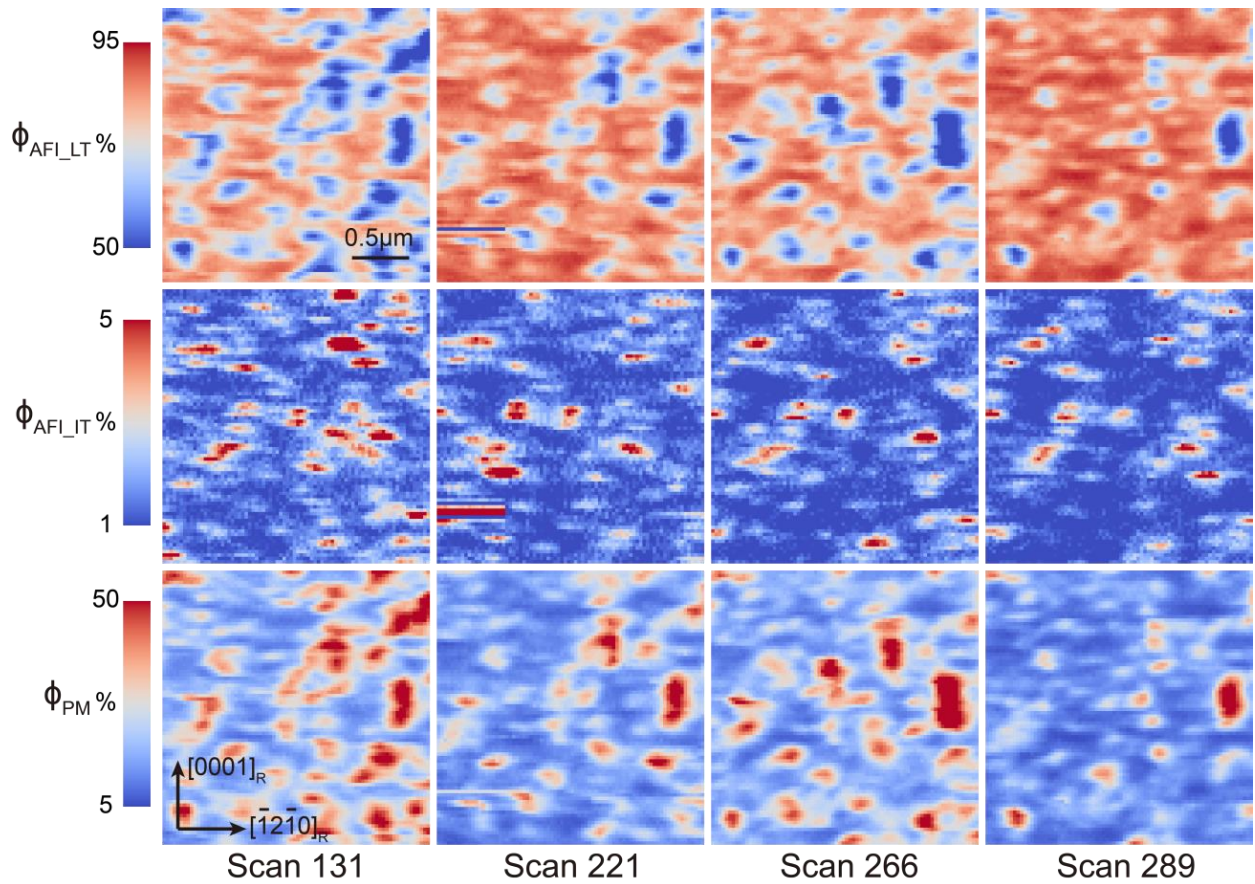

Spatial distribution of the phases in $V_2O_3$, $\Phi_P(i,j)$, where $P = PM, AFI_{LT}, AFI_{IT}$, at 173 K in multiple heating cycles. The morphology of the phases distribution in different heating cycles is largely repeatable.

**Figure S5**

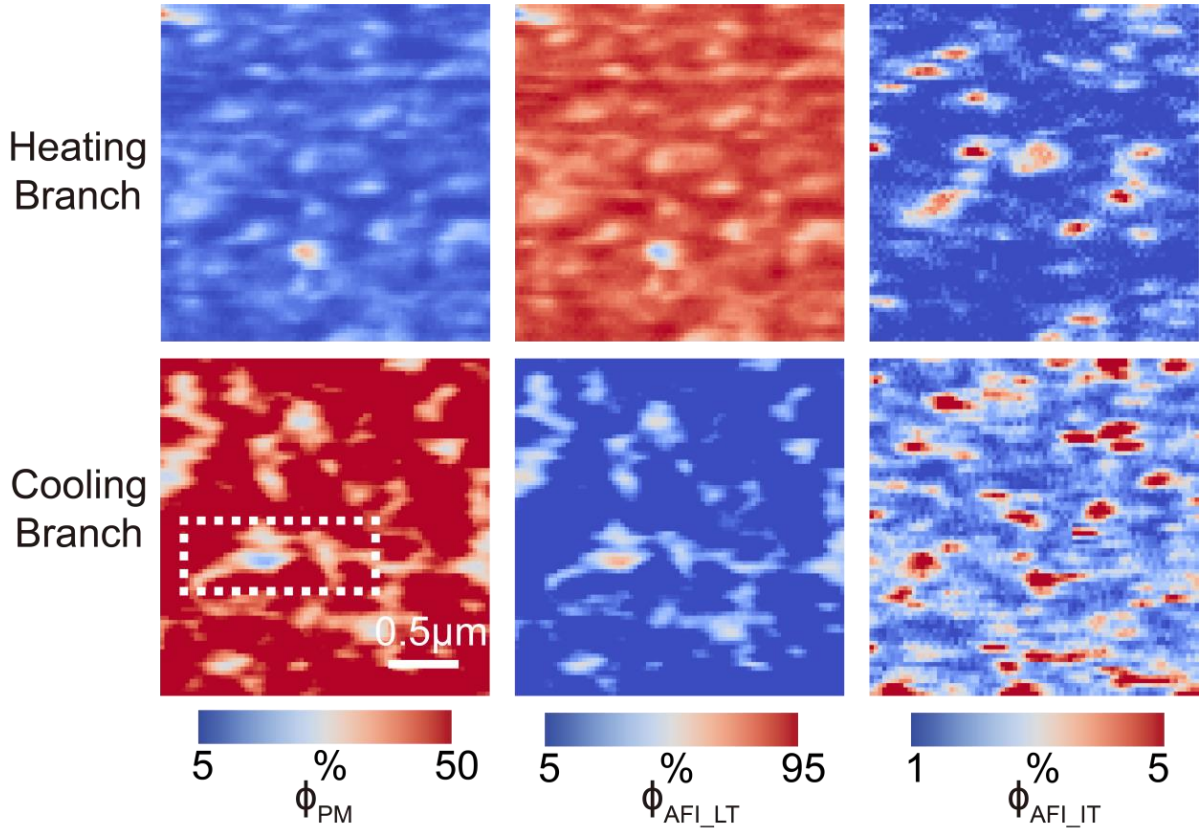

Spatial distribution of the phases in $V_2O_3$, $\Phi_P(i,j)$, where $P = PM, AFI_{LT}, AFI_{IT}$, at 170 K in both heating and cooling branch of the same temperature cycle (Scan 283 and Scan 331, see Fig. S2). At 170 K, the volume of the PM ($AFI_{LT}$) phase is much larger (smaller) in the cooling branch than the heating branch consistent with the temperature hysteresis observation in main text Figure 1C and 1D. Similar spatial distribution of the phases during cooling and heating are observed by comparing the white box in the cooling branch and those in Figure 2. This similarity suggests the structural transition hysteresis is present locally at the nanoscale. Also, different nucleation positions are observed for the PM phase and the $AFI_{LT}$ phase, likely due to the strain heterogeneity and the arrangement of the pre-existing domains. Moreover, comparing the two $AFI_{IT}$ maps, a larger $AFI_{IT}$ volume is observed during the cooling branch, agreeing with the parallel beam observation shown in the inset of Figure 1C.

**Figure S6**

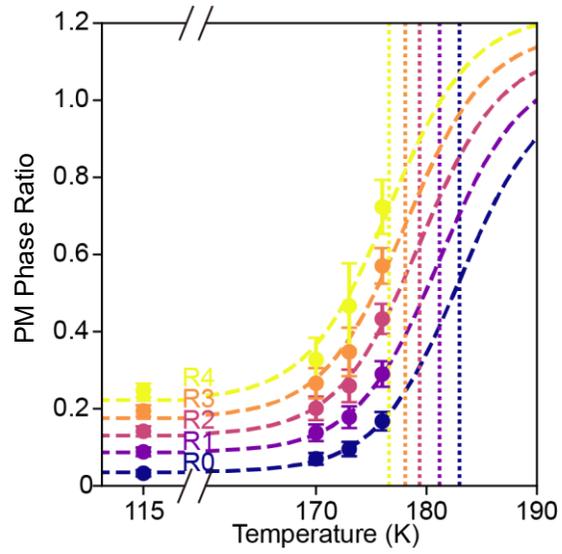

Fitting the temperature dependence of the PM phase fraction in each region with the logistic function, $\frac{1}{e^{-A(T-T_{SPT})}+1}$. The parameter A=0.27 is determined by fitting the curve of the PM phase ratio in Figure 1C up to 175 K. The data and fitted curves (dashed line) are shifted vertically for clarity. The vertical dotted lines mark the transition temperature, $T_{SPT}$, for each region. The transition temperature increases from 176 K in R4 to 183 K in R0.

**Figure S7**

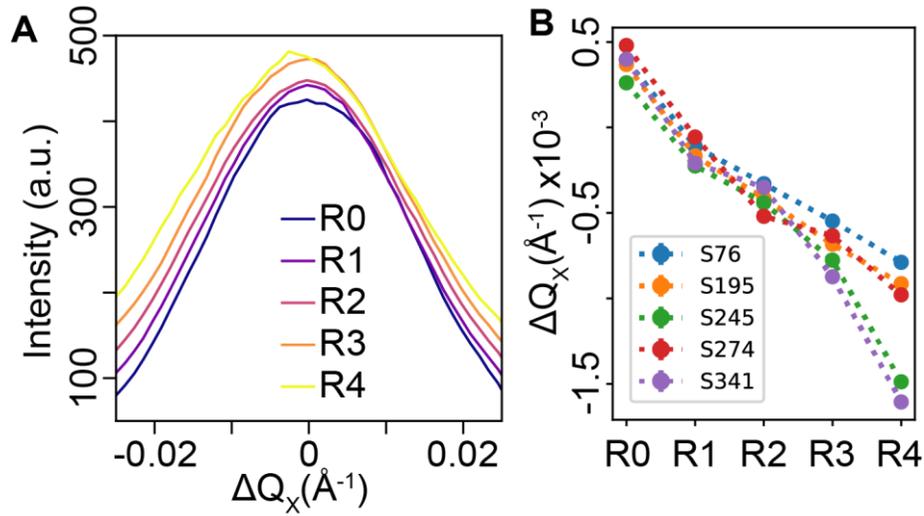

**(A)** Bragg intensity profiles of the $2\bar{3}1_M$, $231_M$ (AFI$_{LT}$) peaks at 115 K projected along the $Q_x$ direction. For each profile, only the detector images measured on the region belonging to the specific interval are summed (The curves are shifted vertically for clarity). By comparing the profiles of different regions, we found a consistent shifting of the peaks: the region with larger $\bar{\Phi}$ values will shift towards the negative direction of $Q_x$ and vice versa. **(B)** The same peak shifting is observed in the 115 K scans of multiple heating cycle. The $\Delta Q_x$ positions are gained through fitting the corresponding intensity profile with gaussian function.

**Figure S8**

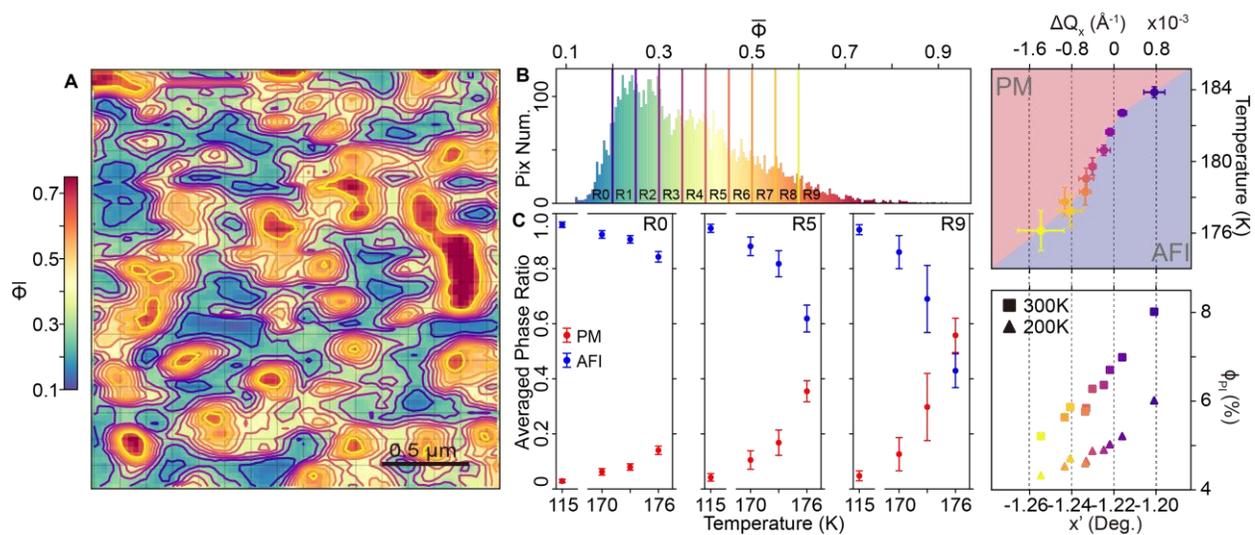

Duplicate of main text Figure 3 analysis with ten equidistant intervals instead of five. As shown in the figure, the results presented in the main text are insensitive to the intervals selected.